\documentclass[onecolumn,showpacs]{revtex4-1}
\usepackage{amsmath}
\usepackage{epsfig}
\usepackage{graphicx}

\begin{document}

\title{Specific heat of underdoped cuprate superconductors from \\ a phenomenological layered Boson-Fermion model}
\author{P. Salas}
\author{ M. Fortes}
\author{M. A. Sol\'{\i}s}
\author{ F. J. Sevilla}
\affiliation{Instituto de F\'{\i}sica, Apartado postal 20-364, Universidad Nacional Aut\'onoma de M\'exico, \\ 01000 M\'exico, D.F., MEXICO}
\date{ \today }

\begin{abstract}
We adapt the Boson-Fermion superconductivity model to include layered systems such as underdoped cuprate superconductors. 
These systems are represented by an infinite layered structure containing a mixture of paired and unpaired fermions. 
The former, which stand for the superconducting carriers, are considered as noninteracting zero spin composite-bosons with a linear energy-momentum dispersion relation in the CuO$_2$ planes where superconduction is predominant, coexisting with the unpaired fermions in a pattern of stacked slabs.
The inter-slab, penetrable, infinite planes are generated by a Dirac comb potential, while paired and unpaired electrons (or holes) are free to move parallel to the planes. 
Composite-bosons condense at a critical temperature at which they exhibit a jump in their specific heat. 
These two values are assumed to be equal to the superconducting critical temperature $T_c$ and the specific heat jump reported for YBa$_{2}$Cu$_{3}$O$_{6.80}$ to fix our model parameters namely, the plane impenetrability and the fraction of superconducting charge carriers. 
We then calculate  the isochoric and isobaric electronic specific heats for temperatures lower than $T_c$  of both, the composite-bosons and the unpaired fermions, which matches recent experimental curves. 
From the latter, we extract the linear coefficient  ($\gamma_n$) at $T_c$, as well as the quadratic ($\alpha T^2$) term for low temperatures. 
We also calculate the lattice specific heat from the ARPES phonon spectrum, and add it to the electronic  part, reproducing  the experimental total specific heat at and below $T_c$ within a $5 \%$ error range, from which the cubic ($\ss T^3$) term for low temperatures is obtained.
In addition, we show that this model reproduces the cuprates mass anisotropies.

\end{abstract}

\keywords{Underdoped cuprate superconductors, critical temperature, specific heat,
Bose-Einstein condensation, multilayers, ... }
\pacs{74.20.-z, 74.25.-q, 74.72.kf}
\maketitle


\section{Introduction}

Since the discovery of cuprate High Temperature Superconductors (HTSC)\cite{Bednorz} in 1986 there has been an extraordinary theoretical effort to explain the nature of their microscopic behavior as they are not completely described by the BCS theory \cite{BCS}. However, very few of these theories consider comparison with specific heat data (or other thermodynamic properties).
The HTSC cuprates were the most frequently studied both experimentally and theoretically until Fe based superconductors showed up \cite{Chubukov2015,Johnston2010,Luo2015}. 
More recently, the appearance of H$_2$S at high  pressures \cite{Drozdov2015,Hirsh2015} beat down the record of higher $T_c$ held by the cuprates.
But, as can be seen in the many publications related to these newer  materials (Fe-based and H$_2$S), the scenario has become even more entangled.
 
High $T_c$ cuprate superconductors have peculiarities  which represent a benchmark in our current understanding of superconductivity. 
It is widely accepted that the Cooper pairs, which are responsible for the superconductivity, move in the copper oxide planes resembling a quasi-2D layered system \cite{Poole2007}, and have coherence lengths much smaller than those in conventional superconductors. 
The phase diagram \cite{Shekhter2013} of the YBa$_{2}$Cu$_{3}$O$_{x}$   shows a dome in the superconducting temperature as a function of the oxygen content $x$, either by introducing electrons or holes, giving the latter ones the higher temperatures.
Cooper pairs are pre-formed at a  particular temperature $T^{\ast} > T_c$ (pseudogap temperature)  above the superconducting dome in the underdoped region \cite{Shekhter2013,Cooper2014} (where $T_c$ is smaller than the higher $T_c$ possible), and they undergo a Bose-Einstein Condensation (BEC) as temperature is lowered \cite{Casas2001,Eagles}. 
There are many other characteristics, but of special interest to us are some recently reported experimental results: the modification of the size of the lattice with oxygen doping \cite{Liang2006}; the notorious increase of the superconducting gap magnitude \cite{Sutherland2003} and the dramatic drop of the Fermi temperature $T_F$  when  doping is diminished \cite{Sebastian2010}. 
These features are of crucial importance in our results.

Experiments, reveal four key characteristics of specific heat as a function of temperature: a linear term $\gamma _{n}$ in the electronic component, which is  believed to come from the normal state electronic specific heat $C_{en}$ above $T_c$ (see Ref. [\onlinecite{Fisher2007}] and references therein), and its superconducting counterpart that evolves as $\gamma_0$ when the temperature approaches zero. 
Secondly, there is a quadratic $\alpha T^2$ term for zero magnetic field \cite{Wright99,Moler97,WangPhysRevB2001} below $T_c$ (not exhibited in conventional superconductors), which changes to a $H^{1/2}T$ component in the presence of an external magnetic field \cite{Volovik93} $H$, attributable to the superconducting part of the electronic specific heat $C_{es}$. 
The reported values for this two constants depend strongly on the conditions of each experiment \cite{WangPhysRevB2001} and on the theoretical method used to relate the different parts of the total specific heat. 
Thirdly, there is a ``jump'' in the constant pressure specific heat \cite{Junod97} $C_p$ at $T_c$  (at zero magnetic field and evolving into a ``peak'' at finite magnetic field), $\Delta C_p/T_c$, attributable to the $C_{es}$ component, indicating a second order phase transition  which in turn becomes a smooth maximum as doping decreases \cite{Loram98}. 
Finally, as the ``upturn'' in the specific heat  at very low temperatures is suppressed,  a  cubic term $\ss T^3$ is also observed \cite{Moler1994}.

The lattice specific heat $C_{l}$ of a cuprate, which is generally considered as not changing with the onset of superconductivity \cite{WangPhysRevB2001}, turns out to give a crucial contribution to the total specific heat. 
Although a series of indirect methods have been used  to extract the electronic component of the total specific heat \cite{Moler97,Wright99,Fisher2007,Loram93}, we use the Phonon Density of States (PDOS) directly derived from Angle Resolved Photoemission Spectroscopy (ARPES) experiments to calculate the lattice specific heat as shown in Refs. [\onlinecite{Cooper2014}] and [\onlinecite{Meingast09}]. 
We obtain that the electronic specific heat contributes less than  $2 \%$ to the total.

In the framework of the most basic Boson-Fermion model of superconductivity \cite{Casas2001,Friedberg1989,Adhikari2000,Friedberg1991}, we assume that Cooper pairs are composite-spin-zero-bosons with either zero or nonzero center-of-mass momenta (CMM), coexisting with a fermion fluid of the unpaired electrons. 
To include the effect of the layered structure of cuprates in the Boson-Fermion model, we calculate the BEC critical temperature and the thermodynamic properties for a system of non-interacting bosons immersed in a periodic multilayer array \cite{PatyJLTP2010,Salas2010}, simulated by an external Dirac comb potential along the perpendicular direction of the CuO$_2$ planes, while the Cooper pairs are allowed to move freely within the planes, with a linear energy-momentum dispersion relation \cite{Adhikari2000}. 
The fermion counterpart is treated in a similar way \cite{PatyJLTP2014} as the boson gas, subject to the same external potential.  
In this model, we assume that only a small fraction $f$ of the initial $N$ fermions available for pairing  participate in the superconductivity at temperature $T=T_c$ and below, where the number of preformed pairs is large enough to achieve coherence independently of the mechanism by which the pairs are formed. 
This latter assumption is based on the analysis of Uemura's plot (Fig. 2 of Ref [\onlinecite{Uemura2006}]) that shows that critical temperatures for cuprates are in the empirical range \cite{AdhikariPhysC2000} of $T_c \approx (0.01 - 0.06) T_F$. 
However, we are aware that the number of pairs could increase as the temperature is lowered below $T_c$, but not so much as to drastically modify the final results. At this stage we assume that the number of pairs remains constant. 

This paper is organized as follows. In Sec. \ref{The system} we lay out the model from which we derive all the thermodynamic properties for the mixture of composite-boson gas coexisting but non-interacting with the unpaired fermion gas immersed in the layered system.
This model depends on two physical parameters: the  impenetrability $P_{0}$ of the planes, which is responsible for the mass anisotropy $M/m$ observed in the cuprates, and the fraction $f$ of superconducting carriers able to condense. 

The expressions for the isobaric electronic specific heat, i.e.,  the superconducting $C_{pes}$ from the Cooper pairs  and the normal  $C_{pen}$ from the unpaired fermions are derived in Sec. \ref{Cpelec}. 
The model parameters are unambiguously determined by the phenomenological properties of YBa$_{2}$Cu$_{3}$O$_{6.80}$, namely the experimental $T_{c}$ and the magnitude of jump $\Delta C_p/T_c$, from which the observed $T^2$ behavior at low temperatures of $C_{pes}$ and the linear dependence on $T$ of $C_{pen}$ are directly obtained. 
Furthermore, we combine $C_{pes}+C_{pen}$ to show that the total electronic specific heat coincides with the experimental results \cite{Meingast09}. 
Our electronic specific heat constants, $\gamma_n(T_c)$  and $\alpha$ are found to be of the same order of magnitude as the experimental values \cite{Junod89,Emerson1999}. 

In Sec. \ref{Total} we calculate the  specific heat for the lattice $C_l$ using the phonon spectrum obtained by ARPES experiments, and compare it to the one obtained using the PDOS from inelastic neutron scattering (INS) experiments \cite{Renker88,Arai1992}. 
At the end, we add these three specific heat contributions, $C_{pes}+C_{pen}+C_l \equiv C^T_p$,  and compare the result with raw data from the experiments. 
From $C^T_p$ the cubic coefficient $\ss$ for low temperatures is extracted giving an excellent agreement with experiments \cite{WangPhysRevB2001,Emerson1999,Moler1994}. 

As a bonus, this model allows us to directly relate the plane impenetrability to the mass anisotropy observed in the HTSC cuprates. This is derived in Sec. \ref{Massani}, giving a  prediction for $M/m$ consistent with the values reported by experiments  \cite{Roulin98,Junod99,Chiao2000}.
Finally, in Sec. \ref{conclusions} we present our conclusions.

\section{\label{The system} Layered structure of underdoped YB\lowercase{a}$_{2}$C\lowercase{u}$_{3}$O$_{x}$ cuprates }

We consider $N$ electrons (or holes) of mass $m_e$ confined in a periodic layered array along the $z$-direction, which mimics the crystallographic structure of the cuprates, and free to move in the other two directions. 
The electrons interact via a BCS-type potential, such that when their energies lie within a shell of width $2\hbar \omega_D$ around the Fermi energy $E_F$ of the system, where $\hbar \omega_D$ is the Debye energy, the electrons are able to form {\it pairs} in  momentum space, but only a fraction $f$ of them will become pairs, leaving a set of {\it pairable but unpaired} electrons. In addition, there are {\it non-pairable} electrons outside this shell. 
Based on this model, we will group the $N$ electrons in two major components: Cooper-pairs  (boson gas) formed by a fraction $f$ of half the total $N$  electrons inside the pairing shell, and a group of ($1-f$) electrons (fermion gas) consisting of the pairable plus the unpairable electrons \cite{Casas2001}.

\subsection{\label{bosones}Composite-bosons: Cooper pairs }

We assume the boson-fermion model where the bosons are Cooper-like pairs that appear as resonances in two electrons or two holes as proposed by Friedberg and Lee \cite{Friedberg1989,Friedberg1991}. In our model, there are $N_B$ = $f N/2$ composite-bosons of mass $m$ = $2m_e$. The Hamiltonian is

\begin{equation}
H =\sum_{\mathbf{k},s} \varepsilon _{\mathbf{k}}  a^{\dagger}_{\mathbf{k},s} a_{\mathbf{k},s} +  \sum_{\mathbf{K}} \varepsilon_{\mathbf{K}} b^{\dagger}_{\mathbf{K}} b_{\mathbf{K}} + H_1, \label{H}
\end{equation}
where $a^{\dagger}_{\mathbf{k},s}$ and $b^{\dagger}_{\mathbf{K}}$ are fermion and composite-boson creation operators, respectively, $s$ is the spin and
\begin{equation}
H_1 = \frac{{\mathsf{G}}}{\sqrt{L^3}} [ a_{\mathbf{K}/2+\mathbf{k},s} a_{\mathbf{K}/2-\mathbf{k},s} b^{\dagger}_{\mathbf{K}} v(k) + h. c.]
\end{equation}
is the interaction Hamiltonian that creates/destroys composite bosons from/into two fermions. Here, $\mathbf{K} = (K_{x},K_{y},K_{z}) \equiv \mathbf{k_1} + \mathbf{k_2}$ is the CMM of the pair, $\mathbf{k} \equiv (\mathbf{k_1} - \mathbf{k_2})/2$ is the relative momentum, $\mathbf{k_1}$ and $\mathbf{k_2}$ the wave vectors of each electron of the pair, and $L^3$ is the volume. 
The form factor $ v(k)$ is normalized such that $ v(0) = 1$, which defines the coupling constant $\mathsf{G}$. 
In our model, we assume the zeroth-order approximation \cite{Friedberg1989,Friedberg1991} so that we keep a mixture of two independent particle systems in a layered structure. 
The solutions of the Schr\"{o}dinger equation associated to the Hamiltonian (\ref{H}), without the $H_1$ term, may be separated in the $x - y$ and $z -$directions, so that the energy for each boson particle  is $\varepsilon_{K} = \varepsilon _{K_{x,y}} + \varepsilon _{K_{z}}$, where $\varepsilon _{K_{x,y}} \equiv 2E_F - \Delta_K$ is the energy of the pair in the $a-b$ crystallographic plane, with $\Delta_K$ the temperature independent binding energy. 

For  $\mathbf{K} \neq 0$, the energy from the Cooper equation may be expanded in a series of powers \cite{Adhikari2000} where the linear term predominates
\begin{equation}
\varepsilon _{K_{x,y}} = \mathsf{e}_0 + C_{1}(K_{x}^{2}+K_{y}^{2})^{1/2},  \label{Green0}
\end{equation}%
with $\mathsf{e}_0 \equiv 2E_F - \Delta_0$ a constant depending on the BCS energy gap $\Delta_0 = 2 \hbar \omega_D \exp(-1/\lambda)$ at $\mathbf{K} = 0$ and $T=0$, $C_{1}=(2/\pi )\hbar \mathsf{v}_{F2D}$ is the linear term coefficient in 2D,  $\mathsf{v}_{F2D}$ is the Fermi velocity also in 2D, $\lambda \equiv g(E_F)V$ is the dimensionless coupling constant in terms of the electronic density of states at the Fermi sea $g(E_F)$ and $V$, the non-local interaction between fermions. 

Along the $z$-direction we use the Kronig-Penney  potential in the  Dirac delta limit, following the scheme we previously developed for a boson gas inside a layered structure \cite{PatyJLTP2010,Salas2010}. The energies are implicitly obtained from the transcendental equation 
\begin{equation}
P_{0}(a/\lambda _{0})\sin (\alpha_{K_{z}} a)/\alpha_{K_{z}} a+
\cos (\alpha_{K_{z}} a)=\cos (K_{z}a),
\label{KPdelta}
\end{equation}
with $\alpha_{K_{z}} ^{2}\equiv 2m\varepsilon _{K_{z}}/\hbar ^{2}$ and $P_{0} = m\Lambda\lambda _{0}/\hbar ^{2}$ is a dimensionless parameter which is a measure of the {\it plane impenetrability} . 
The constant $\lambda _{0} \equiv h/\sqrt{2\pi mk_{B}T_{0}}$ is the de Broglie thermal wavelength of an ideal boson gas in an infinite box at the BEC critical temperature $T_{0} = 2\pi \hbar ^{2}n_{B}^{2/3}/mk_{B} \zeta (3/2)^{2/3}\simeq 3.31\hbar ^{2}n_{B}^{2/3}/mk_{B}$, with $n_{B}\equiv N_B/(L^{3})$ the boson particle number density and $\Lambda$ is the strength of the delta potentials {$\sum_{n_z=-\infty }^{\infty } \Lambda \delta (z - n_za)$}.

The thermodynamic properties of a boson gas can be derived from the grand potential \cite{Path}   
\begin{gather}
\Omega (T,L^3,\mu )=U - TS-\mu N_{B}  
=\Omega _{0} + k_{B}T\sum_{\mathbf{K}{\neq 0}}\ln \bigl\{1-\notag \\ 
\exp [-\beta (\mathsf{e}_0 + C_{1}(K_{x}^{2}+K_{y}^{2})^{1/2}+ 
\varepsilon _{K_{z}}- \mu )] \bigr\},  
\label{omega}
\end{gather}
where $U$ is the internal energy, $S$ the entropy, $\mu $ the boson chemical potential, $\beta \equiv 1/k_{B}T$, and the first term in the rhs corresponds to the $\mathbf{K}={0}$ ground state  contribution $\Omega _{0}$ = $k_{B}T\ln \{1-\exp [-\beta (\varepsilon _0 + \mathsf{e}_0-\mu )]\} $, with  $\varepsilon _{0}  \equiv \hbar ^{2}{\alpha_{0}}^{2}/2m$ the solution of Ec. (\ref{KPdelta}) for the ground state energy.

Expanding the logarithmic function, substituting sums by integrals in the thermodynamic limit, and evaluating the $x,y$ integrals one obtains
\begin{gather}
\Omega \left( T,L^3,\mu \right)=k_{B}T\ln \bigl\{1- \exp[-\beta(\varepsilon _{0} 
+ \mathsf{e}_{0} -\mu)\bigr\}- \notag  \\
\frac{L^{3}}{\left( 2\pi \right) ^{2}}
\frac{\Gamma(2)}{C_{1}^{2}}\frac{1}{\beta ^{3}} {\int_{-\infty }^{\infty }dK_{z}} \ 
\mathsf{g}_{3} (z_b),  
\label{GPpairs}
\end{gather}
where we have used the Bose functions \cite{Path} $\mathsf{g}_{\sigma }(t) \equiv \sum_{l=1}^{\infty}(t)^{\mathit{l}}/\mathit{l}^{\sigma}$  and defined  $z_b \equiv \exp[-\beta (\varepsilon_{K_{z}} + \mathsf{e}_{0} - \mu )] $. 

\subsection{\label{Ferm}Normal state electrons}

The unpaired electrons have the grand potential for an ideal Fermi gas immersed in a layered structure  \cite{PatyJLTP2014} 
\begin{equation}
\Omega (T,L^{3},\mu_{F} )=-k_{B}T\sum_{k= 0}\ln \bigl\{1+\exp [-\beta(\varepsilon _{k}-\mu_{F} )]\bigr\},  \label{omegaFer}
\end{equation}
where $\mu_{F}$ is the chemical potential of the electron gas and $\varepsilon _{k} = \hbar^2 k_x^2/2m_e + \hbar^2 k_y^2/2m_e + \varepsilon_{k_z}$ is the energy of each electron free in the $x-y$ directions and constrained by the permeable planes in $z$-direction. 
As we did in the case of the boson gas, the energy $\varepsilon_{k_z}$ comes from the KP Eq. (\ref{KPdelta}), where we replace $K_z$ by $k_z$ and $P_{0F} = P_0/2$. 
Converting sums to integrals and evaluating the ${x}$, ${y}$ integrals we have 
\begin{gather}
\Omega \left( T,L^3,\mu _{F}\right) =-2\frac{L^{3}}{\left( 2\pi \right) ^{2}}
\frac{m_{e}}{\hbar ^{2}}\frac{1}{\beta ^{2}} {\int_{-\infty }^{\infty }dk_{z}}\mathsf{f}_{2}(z_e),  \label{TGPferm}
\end{gather}
where we use of the Fermi-Dirac functions \cite{Path} $\mathsf{f}_{\sigma }(t)\equiv \sum_{l=1}^{\infty }(-1)^{l-1}t^{\mathit{l}}/ \mathit{l}^{\sigma }$  and  $z_e \equiv \exp [-\beta (\varepsilon _{k_{z}}-\mu _{F})]$. 
From Eq. (\ref{TGPferm}) each thermodynamic property for the fermion gas may be derived.

\subsection{ \label{Crit Temp}Critical temperature}

\begin{figure}[tbh]
	\hspace{-.5cm}
	\epsfig{file=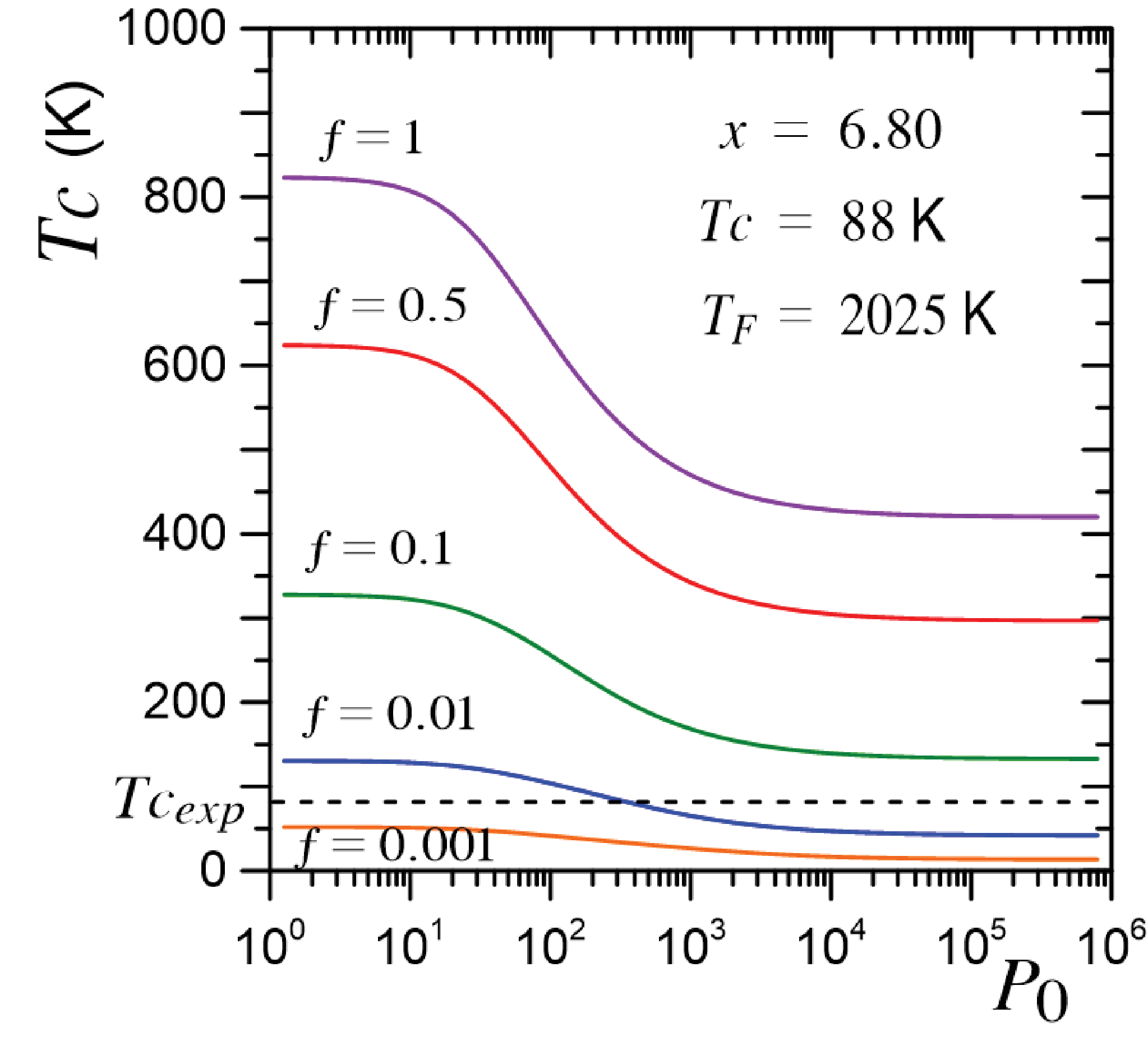,height=9.cm,width=9.cm}
	\vspace{0.cm}
	\caption{ (Color online) Critical temperature as a function of $P_{0}$ for different values of the fraction $f$ of pairable fermions. Dashed line is the experimental $T_{c} = 88$ K for YBa$_{2}$Cu$_{3}$O$_{6.80}$ from Ref. [\onlinecite{Liang2006}].}
	\label{fig:TcvsP0}
\end{figure}

The critical temperature of the cuprate is extracted from the bosonic particle number derived from the grand potential, Eq. (\ref{GPpairs}), namely
\begin{gather}
N_B=\frac{1}{\exp \bigl\{\beta (\varepsilon _{0}+\mathsf{e}_{0}-\mu )
\bigr\}-1} + \notag \\
\frac{L^{3}}{\left( 2\pi\right) ^{2}}\frac{1}
{C_{1}^{2}\beta ^{2}}{\int_{-\infty}^{\infty }dK_{z}} \mathsf{g}_{2}( z_b), \label{Numlin}
\end{gather}%
where the first term on the rhs is the order parameter, i.e., the number $N_{B0}$ of particles in the condensed state and the second term is the number of particles in the excited states.

Setting $T=T_{c}$ in Eq. (\ref{Numlin}) and taking the chemical potential $\mu_{0} = \varepsilon_0 + \mathsf{e}_{0}$ which corresponds to the ground state, so $N_{B0}(T_{c})/N_B\simeq 0$, we have 
\begin{gather}
N_B=
\frac{L^{3}}{\left( 2\pi\right) ^{2}}\frac{1}
{C_{1}^{2}\beta ^{2}}{\int_{-\infty}^{\infty }dK_{z}} \mathsf{g}_{2}\bigl\{ \exp 
[-\beta (\varepsilon _{K_{z}} - \varepsilon _{0} )]\bigr\}, \label{Tc}
\end{gather}%
which must be solved numerically using the fact that in the YBa$_{2}$Cu$_{3}$O$_{x}$ systems there are two copper-oxide regions per unit cell where superconductivity takes place, so the parameter $a$ is fixed to half the crystallographic constant $c$. 

Here we point out the following fact: from the relation of the Fermi energy in terms of the superconducting carrier density \cite{Path}, $E_{F} = {\hbar^2}(3\pi^2)^{2/3} n_s^{2/3}/2m$,
for YBa$_{2}$Cu$_{3}$O$_{6.80}$ with $T_{F}=2025$ K obtained by extrapolation of the $T_F$ curve of Fig. 4 of Ref.  [\onlinecite{Sebastian2010}], we obtain the carrier density as $n_{s} = 9.37\times 10^{26}/$m$^{3}$.
On the other hand, the BEC critical temperature for an ideal Bose gas created from a fermion gas where all fermions are paired \cite{Sevilla}, is $T_{0} = 0.218 T_F$ with $T_F$ the Fermi temperature of the original Fermi gas, so $T_{0}$ gives 441.5 K for this particular superconductor. 
Introducing this value in the definition of $T_0$ given in Sec. \ref{bosones}, one has $n_{B}=1.658\times 10^{26}/$m$^{3}$ for the boson density number, which is almost an order of magnitude smaller than the $n_{s}$ calculated above. 
However, as we mentioned before, by analyzing the Uemura data in Fig. 2 of Ref. [\onlinecite{Uemura2006}] and localizing the diagonal lines labeled as $T = T_F$ and $T = T_0$ (labeled as $T_B$), one would expect that the actual number of superconducting carriers for the cuprates (which we will call $n_b$)  would be about two orders of magnitude smaller than the $n_s$ calculated above. 
Therefore, we assume that only a fraction $f$ of the maximum possible value $n_{B}$ is participating in the boson gas responsible for the superconductivity, hence $n_{b} = fn_{B}$, and we expect this fraction to be $f < 0.01$, as shown in Fig. \ref{fig:TcvsP0}, consistent with the analysis of Refs. [\onlinecite{WangPhysRevB2001}] and  [\onlinecite{AdhikariPhysC2000}]. 

Therefore, the BEC critical temperature for a fraction $f$ of $N_B$ bosons is 
\begin{equation}
T_{0 f}=\frac{2\pi \hbar ^{2}n_{B}^{2/3}}{mk_{B}\zeta (3/2)^{2/3}}
 =T_{0}f^{2/3},  \label{T0Bf}
\end{equation}
the quotient of the fraction of an ideal gas BEC temperature in terms of its $T_F$ is 
\begin{equation}
\frac{T_{0f}}{T_{F}}=\frac{2\pi f^{2/3}}{(6\pi ^{2})^{2/3}\zeta (3/2)^{2/3}}
= 0.218 f^{2/3},  \label{T0fsTF}
\end{equation}
and the thermal wavelenght is $\lambda_{0f}=h/\sqrt{2\pi m k_{B}T_{0f}}= \lambda_{0}/f^{1/3}$. 
For $f = 1$ we recover the case where all pairable fermions participate in the boson gas. 

Additional experimental parameters of  YBa$_{2}$Cu$_{3}$O$_{6.80}$ that we use in our calculations are: the critical temperature \cite{Liang2006} $T_{c exp}$ = 88 K;  the superconducting parameter \cite{Sutherland2003} $\Delta_0 = 50$ meV; the crystallographic \cite{Liang2006}  $c=11.71$ \AA, giving $a=c/2=5.855$ \AA \ and $a/\lambda_0$ = 0.233. 
Finally, we take the height of the jump $\Delta C/T_c$ $\simeq$ 20 mJ/mol K$^2$ from the data published in Ref. [\onlinecite{Emerson1999}]. 
In addition, we use the relation for the Fermi energy $E_{F3{\text D}}=[(3\pi^{2})^{2/3}/2\pi ]E_{F2{\text D}}$ for a 3D system in terms of the Fermi energy for a 2D system $E_{F2{\text D}}=\frac{1}{2}m_{e}\mathsf{v}_{F2D}^{2}$.

In Fig. \ref{fig:TcvsP0} we show the critical temperature as a function of the parameter $P_{0}$ for several values of $f$. The dashed line represents the experimental critical temperature for YBa$_{2}$Cu$_{3}$O$_{6.80}$. 
As can be seen from this figure, there is only a narrow interval
of values of $f \in [0.005,0.05]$, that fits the experimental condition \cite{Emerson1999} $T_{c}$ = 88 K, which in turn determines a set of values of $P_0$. 
This is consistent with our previous assumption that only a small percentage of the initially pairable fermions actually form pairs.
In order to narrow down the range of both values, we obtain the magnitude of the jump in the electronic specific heat from experiments as shown in the next section.

\section{\label{Cpelec} Electronic specific heat}

The cuprate total electronic specific heat at constant pressure $C_{pe}$ is the specific heat of the gas of  Cooper-pairs plus the specific heat of the gas of electrons, $C_{pe}=C_{pes}+C_{pen}$,  each of which is calculated in this section.

\vspace{0.cm}	
\subsection{\label{SuperconductingCe}Superconducting electronic specific heat}

\begin{figure}[tbh]
		\vspace{-1.cm}
		\hspace{-0.5cm}
	\epsfig{file=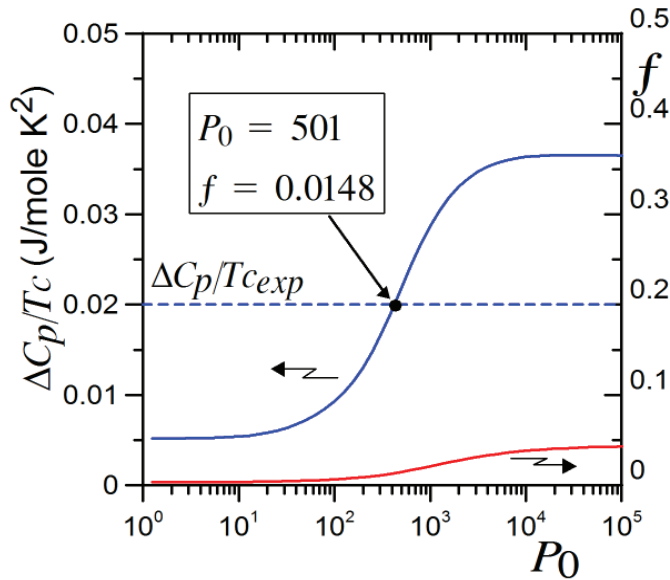,height=11.cm,width=9.cm}
	\vspace{-2.0cm}
	\caption{(Color online) Jump height of the specific heat and $f$ as a function of $P_0$. Dashed line is the experimental $\Delta C_{p}/T_c$ = 20 mJ/mole K$^2$ from Ref. [\onlinecite{Emerson1999}]. \label{fig:DeltaCp} }
\end{figure}

The superconducting electronic specific heat  at constant volume  $C_{Ves}$ for the Cooper pairs gas is $C_{Ves}=\left[ T {\frac{\partial S}{\partial T}}\right] _{N,L^3}$. Hence, taking the fraction $f$, we have 
\begin{widetext}
\begin{gather}
\frac{C_{Ves}}{N_Bk_{B}} =\frac{L^{3}}{f N_B\left( 2\pi\right) ^{2}C_{1}^{2}}
 \bigg[ \frac{2}{\beta} \int_{-\infty}^{\infty } adK_{z} \mathsf{g}_{2}( z_b)
 \Big[ 2\varepsilon_{K_{z}}-\varepsilon_0+\mathsf{e}_0-
 \mu+T\frac{d\mu}{dT}\Big] - \notag  \\
  \int_{-\infty}^{\infty }adK_{z}(\varepsilon_{K_{z}}-
\varepsilon_0) \ln \{1- z_b\} 
\Big[\varepsilon_{K_{z}}+\mathsf{e}_0-\mu +T\frac{d\mu}{dT}\Big]  
+ \frac{6}{\beta^{2}} \int_{-\infty}^{\infty }adK_{z} \mathsf{g}_{3}(z_b)\bigg],  
\label{Cvlinf}
\end{gather}
\end{widetext}
where $\mu$ and its derivative are implicitly obtained from the number equation for  $T \geq T_c$
\begin{gather}
N_B=
\frac{L^{3}}{\left( 2\pi\right) ^{2}}\frac{\Gamma (2)}{C_{1}^{2}}\frac{1}
{\beta ^{2}}{\int_{-\infty}^{\infty }dK_{z}} \mathsf{g}_{2}(z_b ). \label{Numb}
\end{gather}%

The corresponding specific heat at constant pressure is derived from the relation $C_{pes}$ = $C_{Ves}+T L^3 \kappa_T \left[  {\frac{\partial P}{\partial T}}\right]^2 _{N,L^3}$, with $\kappa_T$ the isothermal compressibility. After some algebra, we find
\begin{widetext}
\begin{gather}
\frac{C_{pes}}{N_Bk_{B}} = \frac{C_{Ves}}{N_Bk_{B}} -  
\left( \frac{{\int_{-\infty}^{\infty }dK_{z}} \ln\bigl\{1 - z_b \bigr\}}
{{\int_{-\infty}^{\infty }dK_{z}} \mathsf{g}_{2}(z_b)} \right)  \bigg[ \frac{L^{3}}{f \left( 2\pi \right) ^{2}}
\frac{1}{C_{1}^{2}\beta ^{2}} 
\Big( 3\int_{-\infty}^{\infty }dK_{z} \mathsf{g}_{3}(z_b)   
+ \beta {\int_{-\infty}^{\infty }dK_{z}} \mathsf{g}_{2}(z_b) \Big) \bigg]^2. \label{Cpbos} 
\end{gather}
\end{widetext}
 
In Fig. \ref{fig:DeltaCp} we show the magnitude of the height of the difference between the constant pressure specific heat above and below $T_c$ divided by $T_c$, $\Delta C_{pes}/T_c$, as a function of $P_0$ together with the fraction of Cooper-pairs $f$. 
The horizontal dashed line represents the experimental result \cite{Emerson1999} $|\Delta C_{pes}/T_c|_{exp}$ = 20 mJ/mole K$^2$, and the points where the curves cross this line are the values of $P_0 = 501$ and $f = 0.0148$ that fulfill the conditions for YBa$_{2}$Cu$_{3}$O$_{6.80}$. 
With these two parameters we are able to calculate all the thermodynamic properties for $T \leq T_c$.
 
\subsection{\label{NormalCe}Normal electronic specific heat}

The specific heat at constant volume of the unpaired electrons is obtained from Eq. (\ref{TGPferm})
\begin{widetext}
\begin{gather}
\frac{C_{Ven}}{Nk_{B}} =\frac{1}{(1-f)}\frac{L^{3}}{N \left( 2\pi 
\right) ^{2}}\frac{m_e}{\hbar ^{2}} 
 \bigg[ \frac{1}{\beta } {\int_{-\infty }^{\infty }dk_{z}}\mathsf{f}_{2}(z_e)  \notag  \\
+ 2\int_{-\infty }^{\infty }dk_{z}\ln \{1+ z_e\} 
 \Big[2\varepsilon _{k_{z}}-\mu_{F} +T\frac{d 
\mu_{F}}{d T}\Big]  
+2{\int_{-\infty }^{\infty }dk_{z}}\frac{\varepsilon _{k_{z}}
\{\varepsilon_{k_{z}}-\mu_{F} +T\frac{d \mu_{F} }{d T}\}}
{\exp [\beta (\varepsilon_{k_{z}}-\mu_{F} )]+1} \bigg]. 
\end{gather}
\end{widetext}

Again, the chemical potential $\mu _{F}$  and its derivative are extracted from the corresponding number equation 
\begin{gather}
N=\frac{1}{(1-f)}\frac{2L^{3}}{\left( 2\pi \right) ^{2}}
\frac{m_{e}}{\hbar ^{2}}\frac{1}{\beta }  
 {\int_{-\infty }^{\infty }dk_{z}}\ln \{1+ z_e \}.
\end{gather}

Using the relation for the specific heat at constant pressure, we finally obtain
\begin{widetext}
\begin{gather}
\frac{C_{pen}}{Nk_{B}} = \frac{C_{Ven}}{Nk_{B}} +  
   \frac{2L^{3}}{(1-f)N\left( 2\pi \right) ^{2}}
  \frac{m_{e}}{\hbar ^{2}} \beta \left( \frac{
{\int_{-\infty }^{\infty }}\frac{dk_{z}}
{\exp [\beta (\varepsilon_{k_{z}}-\mu_{F} )]+1}}
{ ( {\int_{-\infty }^{\infty }dk_{z}}\ln \{1+ z_e\} )^2 }  \right) \notag \\
\times \bigg[ 
\int_{-\infty }^{\infty }dk_{z}\ln \{1+ z_e\}\Big[\varepsilon _{k_{z}}-\mu_{F}  +T\frac{d 
	\mu_{F}}{d T}\Big]
+ \frac{2}{\beta} {\int_{-\infty }^{\infty } dk_{z}}\mathsf{f}_{2}(z_e) \bigg]^2. \label{Cpfer} 
\end{gather}
\end{widetext}

\subsection{\label{TotalCe}Total electronic specific heat}

\begin{figure}[tbh]
	\begin{center}
		\vspace{-1.cm}
		\hspace{-0.5cm}
	\epsfig{file=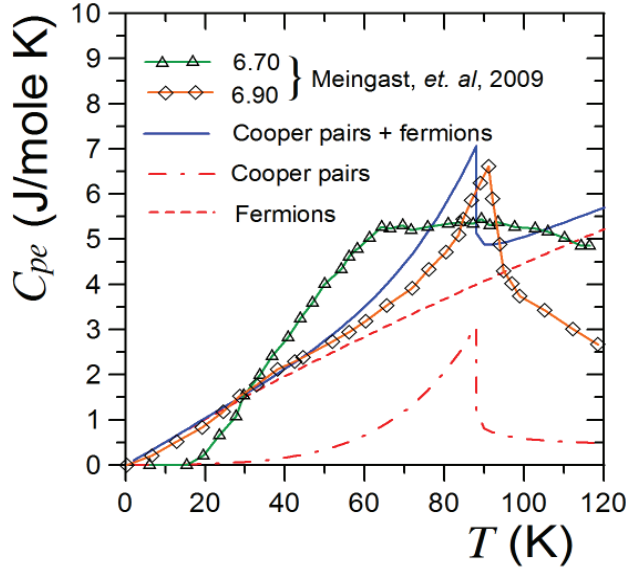,height=11.cm,width=9.cm}
	\end{center}
	\vspace{-2.0cm}
	\caption{(Color online) The Cooper pair and the fermionic  contributions to constant pressure electronic specific heat for  YBa$_{2}$Cu$_{3}$O$_{6.80}$, using  $P_0 = 501$ and $f = 0.0148$, compared to the experimental data of Meingast, {\it et. al.} for two different doping values \cite{Meingast09}.}
	\label{fig:CelectronicosT}
\end{figure}

In Fig.\ref{fig:CelectronicosT} we show the total $C_{pe}$ (continuous line), together with the  Cooper-pairs (dash-dot line) and fermions (dashed line) specific heats. 
We include two experimental curves for YBa$_{2}$Cu$_{3}$O$_{6.70}$ (triangles) and YBa$_{2}$Cu$_{3}$O$_{6.90}$ (diamonds) from Fig. 5 of Ref. [\onlinecite{Meingast09}], where the authors  present exclusively the electronic part after successfully extracting the phonon contribution.

We  obtain the parameter $\gamma_n(T_c) \equiv C_{pen}(T_c)/T_c = 45$ mJ/mol K$^2$ from the linear behavior of the normal electronic specific heat calculated data and  the quadratic term coefficient from the superconducting electronic specific heat $\alpha= 0.038$ mJ/mol K$^3$, which comes from the Cooper pairs. 
Experimental data for $\gamma_n(T_c)$ yields, for example, 32 mJ/mol K$^2$ for $x=0.95$ from Ref. [\onlinecite{Junod89}],  while for  $\alpha$ we find  0.064 mJ/mol K$^3$ for $x=0.80$ from Ref. [\onlinecite{Emerson1999}], both in the same order of magnitude as our results.

There is a \textit{non-zero} value for $\gamma _{0}$, at $T=0$, as stated in Refs. [\onlinecite{Fisher2007,Liang,UherCardwell}]  for oxygen content $x>0.6$, which is different from the extrapolation of $\gamma_n(T_c)$ when $T\rightarrow 0$, suggesting that the pairing mechanism continues to take place.  
However, in our model we are unable to determine $\gamma _{0}$ because we do not include the rate at which pairs continue to form for temperatures below $T_c$ towards $T=0$.  
Furthermore, above $T_c$ paired fermions may decouple through complex mechanisms that we have not considered in this analysis.

Other thermodynamic properties, such as the entropy as well as the Helmholtz free energy of the boson-fermion mixture inside a layered system will be published elsewhere.

\section{\label{Total}Total specific heat}

The total specific heat of YBa$_2$Cu$_3$O$_{6.80}$ is  the electronic plus the lattice specific heat,  i.e., $C_p^T = C_l + C_{pes}+C_{pen}$. In this section we obtain the lattice specific heat  $C_l$ and add it to the electronic contribution, showing that our resulting curves for $C_p^T$ and $C_p^T/T$ lie very close to the raw data reported by some experiments.

\subsection{\label{Lattice}Lattice specific heat}

The total internal energy of a crystal is given by \cite{Kittel}
\begin{equation}
U={\int }\frac{\hbar \omega G(\omega )d\omega }{\left( \exp [\hbar \omega /
	{k_{B}T}]-1\right) },  \label{IntEnerLatt}
\end{equation}%
where  $\omega$ is the vibrational mode frequency, $G(\omega)$ is the PDOS and $\hbar\omega$ is the energy of each mode. The constant volume specific heat for the lattice is then given by 
\begin{equation}
C_{Vl}=k_{B}{\int }\frac{(\hbar \omega /{k_{B}T)}^{2}\exp [\hbar \omega /
	{k_{B}T}]G(\omega )d\omega }{\left( \exp [\hbar \omega /{k_{B}T}]-1\right)
	^{2}}.  \label{SpecHeatLatt}
\end{equation}

\begin{figure}[tbh]
	\begin{center}
		\hspace{-0.5cm}
		\epsfig{file=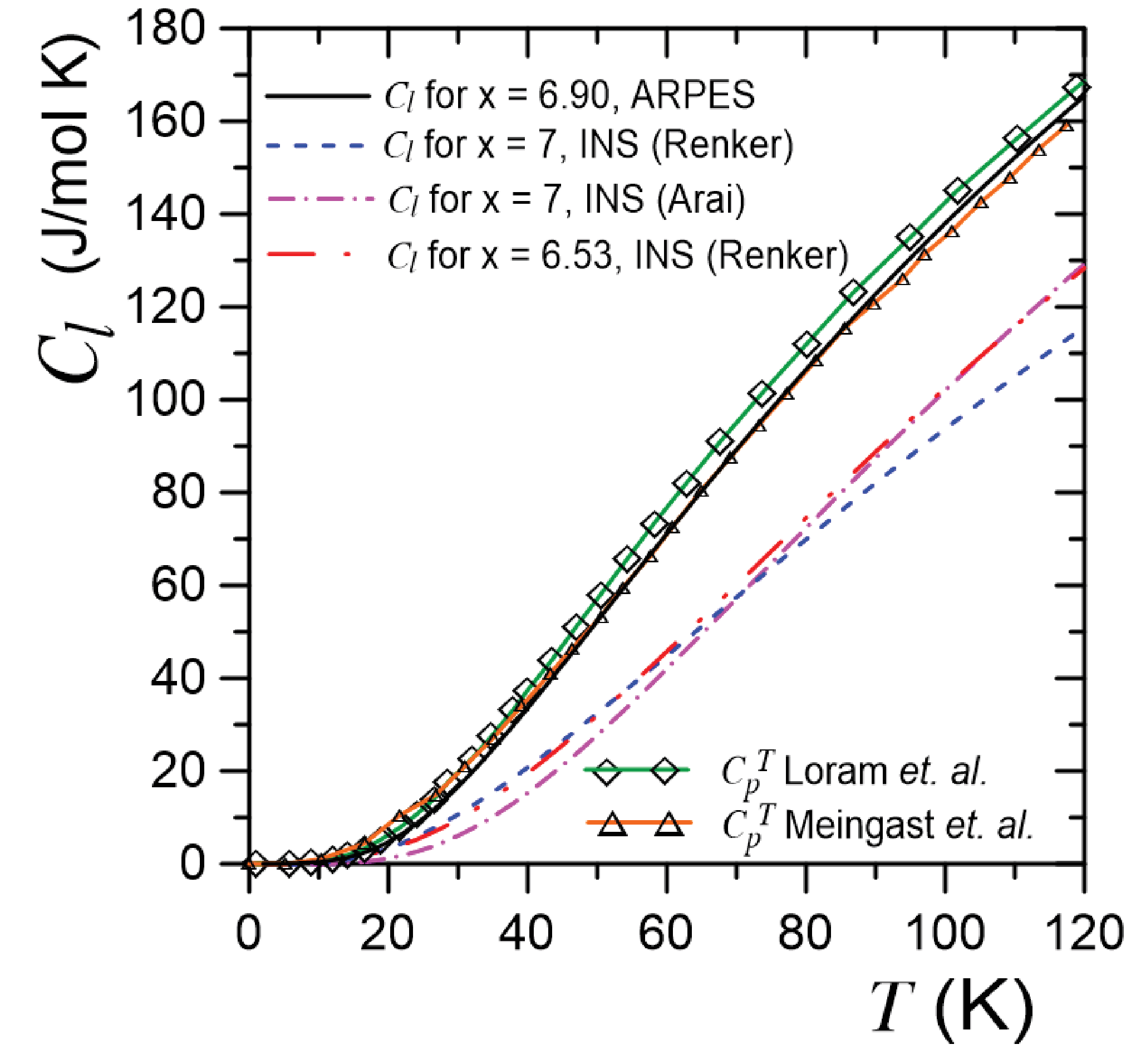,height=9.cm,width=9.cm}
	\end{center}
	\vspace{.0cm}
	\caption{(Color online) Lattice specific heat for different doping values obtained using Eq. \ref{SpecHeatLatt} together with the experimental total specific heat from Refs.  [\onlinecite{Meingast09}] and [\onlinecite{Loram94}].}
	\label{fig:CpLattice}
\end{figure}

We use a phenomenological procedure to calculate the lattice specific heat of the layered cuprate. Specifically, we take the experimental results for the PDOS and introduce it in the theoretical expressions given above.
We analyze the results of three different experiments: two from INS\cite{Renker88,Arai1992}, while the third one is based on a more recent ARPES technique \cite{Meingast09}.
 
Although our lattice specific heat has been calculated at constant volume, a simple calculation of the difference between $C_{pl} - C_{Vl} =  T B V_{mol} \tau^2$ shows that it is smaller than 1 Joule/mol K at $T = T_c$, where $\tau$ is the volumetric thermal expansion coefficient taken from Ref. [\onlinecite{Nagel}], $B$ is the bulk modulus from Ref. [\onlinecite{KoblischkaCardwell}] and $V_{mol}$ is the molar volume. 
Note that this approach already takes into account the anharmonic terms in the lattice component, at least up to the temperature interval considered. 
Therefore we will refer to the lattice specific heat using only the $l$ subindex.

We obtain the results drawn in Fig. \ref{fig:CpLattice} by using the PDOS from the INS  data for YBa$_{2}$Cu$_{3}$O$_{x}$ with $x = 7$  (short dash line)  and $x = 6.53$ (dash-dot-dot line) from the curves of Ref.  [\onlinecite{Renker88}], and for $x =7$ from [\onlinecite{Arai1992}] (dash-dot line), and introduce each one in Eq. (\ref{SpecHeatLatt}) to perform the integrals numerically. 
The difference between the first two curves and the third one is small in the $20$ K $< T <$ $80$ K interval and for $T >$ 80 K it widens progressively. 
In the same Fig. \ref{fig:CpLattice}, we plot our calculation of the lattice specific heat (solid line) using the results from ARPES\cite{Meingast09} for $x = 7$ together with the curves adapted for the total ``raw data" experimental specific heat for $x = 6.67$  from Ref. [\onlinecite{Loram94}] (diamonds) and  for $x = 6.70$ from Ref. [\onlinecite{Meingast09}] (triangles).
We find that our calculated $C_l$ (solid line) using the ARPES density of state is close to the experimental total specific heat $C_p^T$, however, there is a significant difference using INS data (around the 30 $\%$).
	
Three remarks are in order: first, the difference between the lattice specific heat using ARPES and INS around the transition point $T_c$ is at least 30$\%$ (it diminishes as $T$ lowers), which shows that the use of the latter is somehow obsolete and should be discarded, as stated in Ref. [\onlinecite{Cooper2014}]. 
Second, the difference in the lattice specific heat between two near doping values is in general small, and allows us to safely use the $x = 7$ PDOS from ARPES for other dopings, such as $x = 6.80$. 
Finally, in the curves of the experimental $C_p^T$ shown in Fig. \ref{fig:CpLattice}, the jump height is barely noticed, which is another signal that the lattice component is dominant.
For completeness, we calculated the lattice specific heat for the YBa$_{2}$Cu$_{3}$O$_{6}$ non-superconducting compound using the PDOS from INS (not shown in the graphic) which lies very close to the other doping curves obtained also from INS.
Above $T_c$ the lattice specific heat we obtain is still very close to the experimental $C_p^T$ at least for up to $T = 200$ K (not shown).

In summary, the fact that the lattice specific heat $C_l$ and the total experimental specific heat $C_p^T$ are very close leads to the conclusion that the contribution from the electronic components is very small, as will be shown in the next subsection.

\subsection{\label{Sum} Total specific heat of YB\lowercase{a}$_{2}$C\lowercase{u}$_{3}$O$_{6.80}$}

\begin{figure}[tbh]
	\begin{center}
		\vspace{-1.cm}
		\hspace{-.5cm}
	\epsfig{file=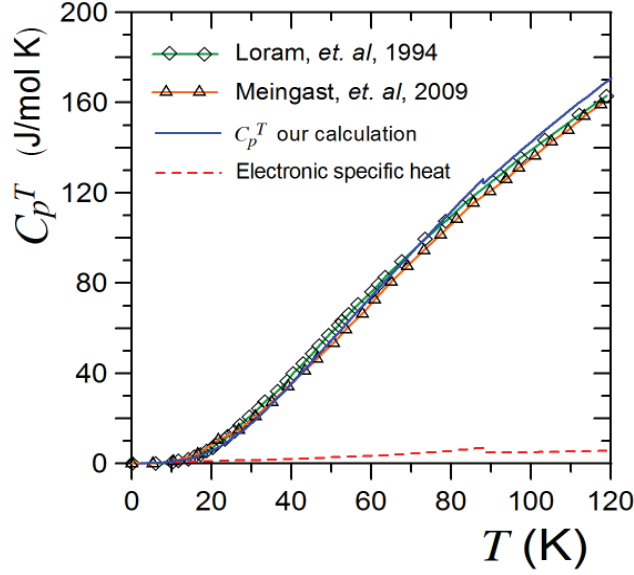,height=11.cm,width=9.cm}
	\end{center}
	\vspace{-2.0cm}
	\caption{(Color online) Total calculated constant pressure specific heat with the results from Refs. [\onlinecite{Meingast09}] and [\onlinecite{Loram94}]. The contribution of the electronic component shown separately.}
	\label{fig:CT}
\end{figure}

\begin{figure}[tbh]
	\begin{center}
		\vspace{-1.cm}
		\hspace{-.5cm}
	\epsfig{file=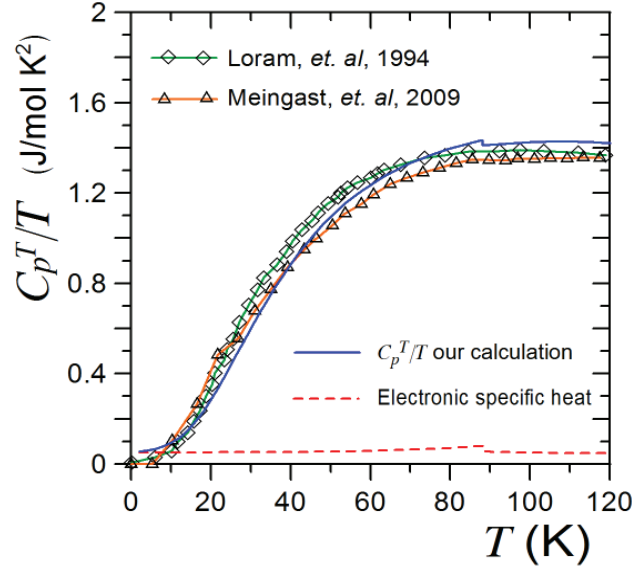,height=11.cm,width=9.cm}
	\end{center}
	\vspace{-2.0cm}
	\caption{(Color on line) Total calculated constant pressure specific heat over temperature with the results from Refs.  [\onlinecite{Meingast09}] and [\onlinecite{Loram94}]. The contribution of the electronic component shown separately.}
	\label{fig:CTsT}
\end{figure}

The total specific heat $C_p^T$ is the sum of all three components: the electronic specific heat we calculated for both  composite-bosons and unpaired electrons with the parameters $P_{0} = 501$ and $f = 0.0148$, in addition to the lattice specific heat from ARPES. 
In Figs. \ref{fig:CT} and  \ref{fig:CTsT} we plot the total specific heat together with the electronic part (normal plus superconducting) to emphasize the size of its contribution. 

In these figures we observe that the total experimental specific heat curves for both, $C_p^T$ and $C_p^T/T$, are satisfactorily reproduced by adding the three analyzed components. 
The minor difference between the experimental shape of the curve and ours for $C_p^T/T$ at and below $T_c$ observed in Fig. \ref{fig:CTsT} may be due to the interactions among the particles and from the dynamical formation of Cooper-pairs for $T < T_c$.

Accordingly, we claim that the contribution of the electronic specific heat to the total is less than the $5 \%$, as suggested in Refs. [\onlinecite{Meingast09}] and [\onlinecite{Loram93}].


Finally, by plotting our total specific heat $C_p^T/T$ vs $T^2$ (not shown), we reproduce the $\ss T^3$ term observed in  experiments  \cite{WangPhysRevB2001,Moler97} for $T <$ 5 K. 
This behavior is expected if the Debye model is used, but it is not trivial for any other lattice model. 
We find  $\ss$ = 0.362 mJ/mol K$^{4}$  compared to 0.333 for $x=6.80$ reported in Ref. [\onlinecite{Emerson1999}], 0.305 for $x=7$ in Ref. [\onlinecite{WangPhysRevB2001}]  and 0.392 for $x=6.50$ in Ref. [\onlinecite{Moler1994}].
 
\section{\label{Massani}Mass anisotropy}

Using the  equations derived in Sec. \ref{bosones} we can make a direct connection between one of the observed features of cuprate superconductors and our model: the mass anisotropy. 

From  Eq. (\ref{KPdelta})  we note that when $P_{0}\rightarrow 0$,  the energy goes to the free-particle energy $\varepsilon_{K_z}\rightarrow \hbar ^{2}K_{z}^{2}/2m$ in the  $z$ direction \cite{Salas2010}. Also, when the particle energies are small, $\varepsilon _{K_{z}} << $ $\hbar ^{2}/2ma^{2}$, one can expand the first term of Eq. (\ref{KPdelta})  around $\varepsilon_0$, so  
\begin{equation}
	\varepsilon _{K _{z}} \cong \varepsilon _{0}+\frac{\hbar ^{2}}{M a^{2}} 	(1-\cos K_{z}a),
	\label{energiapeque}
\end{equation}
where $\varepsilon_0$ satisfies $P_{0}(a/\lambda _{0})\sin (\alpha_{0} a)/\alpha_{0} a+ \cos (\alpha_{0} a)= 1$, and $M $ is the effective mass. 
This last equation is the most commonly used  for quasi-bidimensional models of superconductors \cite{WK}, but it is a model that is constrained only to the first energy band and for zero ground state energy, i.e., $\varepsilon _{0} = 0$ when  ${K_z} = 0$, which is not the general case for layered systems \cite{Salas2010}.

Writing explicitly the effective mass $M$ in Eq. (\ref{energiapeque}) we have 
\begin{equation}
	M/m= | [\sin (\alpha_{0} a)- (P_{0}(a/\lambda_{0})+1) \cos(\alpha_{0} a)/(\alpha_{0} a)]/(\alpha_{0} a) |.
	\label{Mass}
\end{equation}
 
Introducing the values for $P_0, f, a/\lambda_0$ and $\varepsilon _{0}$ previously obtained we get  $M/m$ = 12.3. Experimental reports give 5.3 for $x={7}$ from Refs. [\onlinecite{Junod99,Chiao2000}], 7.0 for $x={6.92}$ from [\onlinecite{Junod99}] and 10.8 for $x=6.80$ from   [\onlinecite{Roulin98}], showing an increasing value dependence as doping lowers, which sets our result within the expected range.

\section{\label{conclusions}Conclusions}

While most procedures take the experimental curves of the total specific heat and subtract components, we qualitatively and quantitatively construct the total constant pressure specific heat for the YBa$_{2}$Cu$_{3}$O$_{x}$ underdoped cuprates from a simple, first principles model: the  Boson-Fermion theory of superconductivity applied to  layered systems. 
The model assumes the Cooper pairs as a composite-boson gas coexisting with an unpaired electrons (or holes) fermion gas. 
Both gases are constrained in a stacked slabs structure modeled by a Dirac comb potential in the perpendicular direction to the CuO$_{2}$ planes. 
Although no residual interactions among Cooper pairs and unpaired fermions are considered, the model reproduces qualitatively and quantitatively the experimental curves of the electronic part and the total specific heat. 

For a specific underdoped cuprate we take the CuO$_2$ plane separation as our $a$ constant. In addition, we use the experimental critical temperature and the electronic specific heat jump  to set our phenomenological parameters: the planes impenetrability $P_0$ and the fraction $f$ of fermions that turn into Cooper pairs. 

The total specific heat is calculated by adding the specific heats coming from the composite-bosons (superconducting electronic specific heat), unpaired fermions (normal electronic specific heat) and the lattice calculated from the ARPES PDOS. 
The resulting curves for YBa$_{2}$Cu$_{3}$O$_{6.80}$ are compared to the experimental results,  giving a remarkable agreement within a $5\%$ error range for temperatures below $T_c$. 

We derive the linear dependence on temperature $\gamma_n T$ and the quadratic one $\alpha T^2$ for the electronic specific heat, obtaining  $\gamma_n(T_c)= 45$ mJ/mol K$^2$ and $\alpha= 0.038$ mJ/mol K$^3$ for $T<T_c$, in agreement with the experimental data reported for similar cuprates, which is an additional check of consistency for our model.
We show that the correspondence relating  the normal electronic specific heat with the unpaired fermions, and the superconducting term with the Cooper pairs, is a valid assumption. 
These  results make plausible the assumption that not all pairable fermions in the Fermi sea are paired, even at temperatures near zero, and that the jump in the specific heat  is a direct consequence of the condensation of the pairs. 
We also confirm that the lattice specific heat from the phonon density of states by ARPES measurements is better than that obtained from INS experiments. 
It can also be seen that the calculated total specific heat shows the same temperature cubic behavior for $T <$ 5 K as shown experimentally, with a coefficient $\ss$ = 0.362 mJ/mol K$^{4}$. 
Additionally, we find that the electronic specific heat (normal plus superconducting ) has a contribution $< 5 \%$ of the total at the transition temperature. 
Finally, another direct outcome is the reproduction of the high mass anisotropy of the cuprates, giving $M/m$ = 12.3 for the compound analyzed. 

The present method may be applied to other HTSC cuprates and to some iron-based superconductors, which will be done in a future publication.

We acknowledge the partial support from grants PAPIIT IN-111613 and  CONACYT 221030.

\end{document}